\documentclass[10pt]{article}
\usepackage{epsfig}
\title{Beam Coupling Impedances of Small Discontinuities\thanks
 {Lectures presented at the US Particle Accelerator School in 
 Tucson, AZ on January 17-21, 2000, as a part of R.L.~Gluckstern's
 course "Analytic Methods for Calculating Coupling Impedances"}} 
\author{Sergey~S.~Kurennoy \\
        SNS Division, Los Alamos National Laboratory,\\ 
        Los Alamos, NM 87545, USA }
\date{}

\begin{document}
\maketitle

\begin{abstract} 
A general derivation of the beam coupling impedances produced by 
small discontinuities on the wall of the vacuum chamber of an 
accelerator is reviewed. A collection of analytical formulas for 
the impedances of small obstacles is presented.
\end{abstract}

\section{Introduction}

A common tendency in design of modern accelerators is to minimize 
beam-chamber coupling impedances to avoid beam instabilities and 
reduce heating. Even contributions from tiny discontinuities
like pumping holes have to be accounted for because of their large 
number. Numerical time-domain methods are rather involved and 
time-consuming for small obstacles, especially for those protruding 
inside the beam pipe. This makes analytical methods for calculating 
the impedances of small discontinuities very important. 

A discontinuity on the wall of a beam pipe --- a hole, cavity,
post, mask, etc. --- is considered to be small when its typical
size is small compared to the size of the chamber cross section 
and the wavelength.
Analytical calculations of the coupling impedances of small obstacles 
are based on the Bethe theory of diffraction of electromagnetic waves 
by a small hole in a metal plane \cite{Bethe}. The method's basic idea 
is that the hole in the frequency range where the wavelength is large 
compared to the typical hole size, can be replaced by two induced 
dipoles, an electric and a magnetic one. The dipole magnitudes are
proportional to the beam fields at the hole location, with the
coefficients called polarizabilities \cite{Collin}. The fields 
diffracted by a hole into the vacuum chamber can be found as those 
radiated by these effective electric and magnetic dipoles, and
integrating the fields gives us the coupling impedances.
Following this path, the impedances produced by a small hole on 
the round pipe were calculated first in \cite{SK92}.

Since essentially the same idea works for any small obstacle, 
the method can be extended for arbitrary small discontinuities
on the pipe with an arbitrary-shaped cross section.
Analytical expressions for the coupling impedances of various 
small discontinuities on the wall of a cylindrical beam pipe with 
an arbitrary single-connected cross section have been obtained in 
Refs.\ \cite{SK92}-\cite{KGS}. 

All the dependence on the discontinuity shape enters only through 
its polarizabilities. Therefore, the problem of calculating the
impedance contribution from a given small discontinuity was 
reduced to finding its electric and magnetic polarizabilities,
which can be done by solving proper electro- or magnetostatic 
problem. Useful analytical results have been obtained for 
various axisymmetric obstacles (cavities and irises) \cite{K&S}, 
as well as for holes and slots: circular \cite{Bethe} and elliptic 
\cite{Collin} hole in a zero-thickness wall, circular \cite{G&D} and 
elliptic \cite{R&G} hole in a thick wall, various slots (some results 
are compiled in \cite{SK93s}), and for a ring-shaped cut \cite{SK96a}. 
The impedances of various protrusions (post, mask, etc) have been
calculated in \cite{SK96}.

This text is organized as follows. In Sections 2 and 3 a general 
derivation of the coupling impedances for a small discontinuity is 
given, and then in Sect.\ 4 the trapped modes are discussed. This 
part mostly follows Ref.\ \cite{KGS}. In Sect.\ 5 we collected for 
the reader convenience some practical formulas for the coupling 
impedances of various small discontinuities. In a compact form,
many of these formulas are included in the handbook 
\cite{AccPhysHandbook}.

\section{Fields}

\subsection{Beam Fields}

Let us consider an infinite cylindrical pipe with an arbitrary 
cross section $S$ and perfectly conducting walls. The $z$ axis is
directed along the pipe axis, a hole (or another small discontinuity
like a post, cavity or mask) is located at the point
($\vec{b},z=0$), and its typical size $h$ satisfies $h\ll b$.
To evaluate the coupling impedance one has to calculate the fields
induced in the chamber by a given current. If an ultrarelativistic 
point charge $q$ moves parallel to the chamber axis with the 
transverse offset $\vec{s}$ from the axis, the fields harmonics 
$\vec{E}^b,\vec{H}^b$ produced by this charge on the chamber wall 
without discontinuity would be
\begin{eqnarray}
 E^b_{\nu} (\vec{s},z;\omega) & = & Z_0H^b_{\tau}(\vec{s},z;\omega)  
 \label{beamf} \\  & = & -Z_0qe^{ikz} \sum_{n,m} k^{-2}_{nm}
 e_{nm}(\vec{s}\,) \nabla _{\nu}e_{nm}(\vec{b}\,)  \ ,  \nonumber       
\end{eqnarray}
where $Z_0 = 120 \pi$~Ohms is the impedance of free space, and 
$k^2_{nm}$, $e_{nm}(\vec{r})$ are eigenvalues and orthonormalized 
eigenfunctions (EFs) of the 2D boundary problem in $S$:
\begin{equation}
\left (\nabla ^2+ k^2_{nm}\right ) e_{nm} =
  0 \ ; \qquad e_{nm}\big\vert_{\partial S} = 0 \ . \label{boundpr}
\end{equation}
Here $\vec{\nabla}$ is the 2D gradient in plane $S$; $k=\omega/c$; 
$\hat{\nu}$ means an outward normal unit vector, $\hat{\tau}$ is 
a unit vector tangent to the boundary $\partial S$ of the chamber 
cross section $S$, and $\{ \hat{\nu},\hat{\tau},\hat{z}\}$ form a 
right-handed basis. The differential operator $\nabla _{\nu}$ is 
the scalar product $\nabla _{\nu}=\hat{\nu} \cdot \vec{\nabla}$. 
The eigenvalues and EFs for circular and rectangular cross 
sections are given in the Appendix.

Let us introduce the following notation for the sum in 
Eq.~(\ref{beamf})
\begin{equation}
 e_\nu(\vec{s}\,) = - \sum_g k^{-2}_g
 e_g(\vec{s}\,) \nabla _{\nu}e_g(\vec{b}\,)  \label{enorm}
\end{equation}
where $g=\{n,m\}$ is a generalized index. From a physical viewpoint,
this is just a normalized electrostatic field produced at the 
hole location by a filament charge displaced from the chamber axis 
by distance $\vec{s}$. It satisfies the normalization condition
\begin{equation}
 \oint_{\partial S}\! dl \ e_\nu(\vec{s}\,) = 1 \ ,  \label{norma}
\end{equation}
where integration goes along the boundary ${\partial S}$, which 
is a consequence of the Gauss law. It follows from the fact that 
Eq.~(\ref{enorm}) gives the boundary value of 
$\vec{e}_\nu(\vec{s}\,) \equiv -\vec{\nabla}
 \Phi(\vec{r}\,-\vec{s}\,)$, where $\Phi(\vec{r}\,-\vec{s}\,)$
is the Green function of boundary problem (\ref{boundpr}):
 $\nabla^2 \, \Phi(\vec{r}\,-\vec{s}\,) = - 
       \delta(\vec{r}\,-\vec{s}\,)$.
For the symmetric case of an on-axis beam in a circular pipe 
of radius $b$ from Eq.~(\ref{norma}) immediately follows 
$e_\nu(0)=1/(2\pi b)$. It can also be derived by directly summing
up the series in Eq.~(\ref{enorm}) for this particular case.

\subsection{Effective Dipoles and Polarizabilities}

At distances $l$ such that $h \ll l \ll b$, the fields radiated by the 
hole into the pipe are equal to those produced by effective dipoles 
\cite{Bethe,Collin} 
\begin{eqnarray}
P_\nu & = & - \chi \varepsilon_0 E^h_\nu/2; \quad  
M_\tau = (\psi_{\tau \tau} H^h_\tau + \psi_{\tau z} H^h_z )/ 2; 
 \nonumber \\
M_z & = & (\psi_{z \tau} H^h_\tau + \psi_{z z} H^h_z )/ 2 
 \ ,                 \label{dip}
\end{eqnarray}
where superscript '$h$' means that the fields are taken at the hole.
Polarizabilities $\psi, \chi$ are related to the effective 
ones $\alpha_e, \alpha_m$ used in \cite{Collin,SK92} as 
$\alpha_e=-\chi/2$ and $\alpha_m=\psi/2$, so that for a circular hole 
of radius $a$ in a thin wall $\psi=8a^3/3$ and $\chi=4a^3/3$ 
\cite{Bethe}. In general, $\psi$ is a symmetric 2D-tensor, 
which can be diagonalized. If the hole is symmetric, and its 
symmetry axis is parallel to $\hat{z}$, the skew terms vanish, 
i.e.\ $\psi_{\tau z}=\psi_{z\tau}=0$. In a more general case of 
a non-zero tilt angle $\alpha $ between the major symmetry axis 
and $\hat{z}$, 
\begin{eqnarray}
 \psi_{\tau \tau} & = & \psi_\bot \cos^2{\alpha} + 
 \psi_{\|} \sin^2{\alpha} \ , \nonumber \\
 \psi_{\tau z} & = & \psi_{z \tau } = (\psi_{\|} - \psi_\bot) 
  \sin{\alpha}\cos{\alpha}  \ , \label{mpol} \\
 \psi_{zz} & = & \psi_\bot \sin^2{\alpha} + \psi_{\|} \cos^2{\alpha} 
  \ , \nonumber
\end{eqnarray}
where $\psi_{\|}$ is the longitudinal magnetic susceptibility (for 
the external magnetic field along the major axis), and 
$\psi_\bot$ is the transverse one (the field is transverse to the 
major axis of the hole).
When the effective dipoles are obtained, e.g., by substituting beam 
fields (\ref{beamf}) into Eqs.~(\ref{dip}), one can calculate the fields 
in the chamber as a sum of waveguide eigenmodes excited in the chamber
by the dipoles, and find the impedance. This approach has been carried
out for a circular pipe in \cite{SK92}, and for an arbitrary chamber 
in \cite{SK92p}. The polarizabilities for various types of small 
discontinuities are discussed in detail in Section 5.

\subsection{Radiated Fields}
 
The radiated fields in the chamber can be expanded in a series in 
TM- and TE-eigenmodes \cite{Collin} as
\begin{eqnarray}
\vec{F} = \sum_{nm}\left [ A^{+}_{nm}\vec{F}^{(E)+}_{nm}\theta (z) +   
 A^{-}_{nm}\vec{F}^{(E)-}_{nm}\theta (-z) \right ] +  \label{fexp} \\
 \sum_{nm}\left [ B^{+}_{nm}\vec{F}^{(H)+}_{nm}\theta (z) +   
 B^{-}_{nm}\vec{F}^{(H)-}_{nm}\theta (-z) \right ] \ , \nonumber
\end{eqnarray}
where $\vec F$ means either $\vec E$ or $\vec H$, superscripts 
'$\pm$' denote waves radiated respectively in the positive (+, $z>0$) 
or negative ($-$, $z<0$) direction, and $\theta (z)$ is the Heaviside 
step function. The fields $F^{(E)}_{nm}$ of $\{n,m\}$th 
TM-eigenmode in Eq.~(\ref{fexp}) are expressed \cite{Collin} 
in terms of EFs (\ref{boundpr})
\begin{eqnarray}
E^\mp_z & = & k_{nm}^2 e_{nm} \exp(\pm \Gamma _{nm}z) \ ; 
     \qquad  H^\mp_z = 0 \ ; \nonumber \\
\vec{E}^\mp_t & = & \pm \Gamma _{nm} \vec{\nabla}e_{nm}
        \exp(\pm \Gamma _{nm}z) \ ;               \label{emode} \\
\vec{H}^\mp_t & = & \frac{ik}{Z_0} \hat{z} \times \vec{\nabla} 
   e_{nm} \exp(\pm \Gamma _{nm}z) \ , \nonumber 
\end{eqnarray}
where propagation factors $\Gamma _{nm}=(k_{nm}^2-k^2)^{1/2}$ should 
be replaced by $-i \beta _{nm}$ with $\beta _{nm}=(k^2-k_{nm}^2)^{1/2}$ 
for $k>k_{nm}$. For given values of dipoles (\ref{dip}) the unknown 
coefficients $A^\pm_{nm}$ can be found \cite{SK92,SK92p} using the 
Lorentz reciprocity theorem 
\begin{equation}
A^{\pm}_{nm} = a_{nm} M_\tau \pm b_{nm} P_\nu \ ,  \label{Apm}
\end{equation}
with 
\begin{equation}
 a_{nm}  = - \frac{ik Z_0}{2 \Gamma _{nm} k_{nm}^2}   
                \nabla_\nu e^h_{nm} \ ;  \quad
 b_{nm} =  \frac{1}{2 \varepsilon_0 k_{nm}^2}  
                \nabla_\nu e^h_{nm} \ .  \label{ab} 
\end{equation}
The fields $F^{(H)}_{nm}$ of the TE${}_{nm}$-eigenmode in 
Eq.~(\ref{fexp}) are 
\begin{eqnarray}
H^\mp_z & = & k'^2_{nm} h_{nm} \exp(\pm \Gamma' _{nm}z) \ ; 
     \qquad  E^\mp_z = 0 \ ; \nonumber \\
\vec{H}^\mp_t & = & \pm \Gamma' _{nm} \vec{\nabla}h_{nm}
        \exp(\pm \Gamma' _{nm}z) \ ;             \label{hmode} \\
\vec{E}^\mp_t & = & -ik Z_0 \hat{z} \times \vec{\nabla} 
   h_{nm} \exp(\pm \Gamma' _{nm}z) \ , \nonumber 
\end{eqnarray}
with propagation factors $\Gamma' _{nm}=(k'^2_{nm}-k^2)^{1/2}$ 
replaced by $-i \beta' _{nm} = -i(k^2-k'^2_{nm})^{1/2}$ when 
$k>k'_{nm}$. Here EFs $h_{nm}$ satisfy the boundary problem 
(\ref{boundpr}) with the Neumann boundary condition 
$\nabla_\nu h_{nm}\vert_{\partial S} = 0$, and $k'^2_{nm}$ 
are corresponding eigenvalues, see in Appendix. The TE-mode 
excitation coefficients in the expansion (\ref{fexp}) for the 
radiated fields are 
\begin{equation}
B^{\pm}_{nm} = \pm c_{nm} M_\tau + d_{nm} P_\nu 
                              +  q_{nm} M_z  \ ,  \label{Bpm}
\end{equation}
where 
\begin{eqnarray}
 c_{nm} & = & \frac{1}{2 k'^2_{nm}} \nabla_\tau h^h_{nm} \ ;  
 \quad q_{nm}  =  \frac{1}{2 \Gamma'_{nm} } h^h_{nm} \ ; 
 \nonumber  \\
 d_{nm} & = & - \frac{ik }{2 Z_0 \varepsilon_0 
 \Gamma'_{nm} k'^2_{nm}} \nabla_\tau h^h_{nm} \ .       \label{cdq} 
\end{eqnarray}

\subsection{Fields near Hole with Radiation Corrections}

A more refined theory should take into account the reaction 
of radiated waves back on the hole. Adding corrections to the 
beam fields (\ref{beamf}) due to the radiated waves in the vicinity 
of the hole gives 
\begin{eqnarray}
E_\nu & = & \frac{ E^b_\nu + \psi_{z\tau} \Sigma'_x Z_0 H_\tau +  
 \psi_{zz} \Sigma'_x Z_0 H_z } { 1 - \chi ( \Sigma_1 - \Sigma'_1 ) } ,  
                                                      \label{En} \\
H_\tau & = & \frac{H^b_\tau + \psi_{\tau z} ( \Sigma_2 - \Sigma'_2 ) 
 H_z} {1 -\psi_{\tau\tau} (\Sigma_2 - \Sigma'_2 )} ,  \label{Ht} \\
H_z & = & \frac { \chi \Sigma'_x E_\nu  /Z_0 + \psi_{z\tau} 
 \Sigma'_3 H_\tau } { 1 - \psi_{zz} \Sigma'_3 }  ,    \label{Hz} 
\end{eqnarray}
where ($s=\{n,m\}$ is a generalized index) 
\begin{eqnarray}
\Sigma_1 & = & \frac{1}{4} 
 \sum_s { \frac{ \Gamma_s \left (\nabla_\nu e^h_s \right)^2} 
  {  k^2_s } }  \ ;  \  
\Sigma_2  =  \frac{k^2}{4} 
 \sum_s { \frac{\left (\nabla_\nu e^h_s \right)^2} 
  { \Gamma_s k^2_s } }  \ ;               \nonumber  \\
\Sigma'_1 & = &  \frac{k^2}{4} 
 \sum_s \frac{ \left (\nabla_\tau h^h_s \right)^2} 
       {\Gamma'_s k'^2_s}  \ ; \ 
\Sigma'_2  =  \frac{1}{4} 
 \sum_s \frac{\Gamma'_s \left (\nabla_\tau h^h_s \right)^2} 
       {k'^2_s}  \ ;              \nonumber   \\
\Sigma'_x & = &  i \frac{k}{4} \sum_s 
\frac{ h^h_s \nabla_\tau h^h_s }{ \Gamma'_s }  \ ; \ 
\Sigma'_3  =  \frac{1}{4} 
 \sum_s \frac{ k'^2_s \left ( h^h_s \right)^2} 
       {\Gamma'_s }  \ .                   \label{sums}   
\end{eqnarray}
Since this consideration works at distances larger than $h$, 
one should  restrict the summation in Eq.~(\ref{sums}) to the 
values of $s=\{n,m\}$ such that $k_s h \le 1$ and $k'_s h \le 1$. 

\section{Impedance}

\subsection{Longitudinal Impedance}

The generalized longitudinal impedance of the hole depends on the 
transverse offsets from the chamber axis $\vec{s}$ of the leading 
particle and $\vec{t}$ of the test particle, and is defined 
 \cite{SKrev} as  
\begin{equation}
Z(k;\vec{s},\vec{t}\:) = - \frac{1}{q} \int _{-\infty}^{\infty} 
dz e^{-ikz} E_z(\vec{t},z;\omega ) \ ,          \label{impdef}
\end{equation}
where the longitudinal field $E_z(\vec{t},z;\omega )$ is taken along 
the test particle path. The displacements from the axis are assumed 
to be small, $s \ll b$ and $t \ll b$. The impedance 
$Z(k;\vec{s},\vec{t}\:)$ includes higher multipole longitudinal 
impedances, and in the limit $s,t \to 0$ gives the usual monopole 
one $Z(k)=Z(k;0,0)$. To calculate $E_z(\vec{t},z;\omega )$, we use 
Eq.~(\ref{fexp}) with coefficients (\ref{Apm}) and (\ref{Bpm}) 
in which the corrected near-hole fields  (\ref{En})-(\ref{Hz}) are 
substituted [a dependence on $\vec{s}$ enters via beam fields 
(\ref{beamf})]. It yields
\begin{eqnarray}
 && Z(k;\vec{s},\vec{t}\:) = - \frac{ikZ_0e_\nu(\vec{s}\,)
 e_\nu(\vec{t}\:)}{2}   \times       \label{imp} \\ && \times \ 
 \left [ \ \frac{\psi_{\tau\tau}}{1 - \psi_{\tau\tau} ( \Sigma_2 - 
 \Sigma'_2 )} + \ \psi^2_{\tau z} \Sigma'_3 - \frac{\chi}
 {1 - \chi ( \Sigma_1 - \Sigma'_1 )} \ \right ] \ , \nonumber
\end{eqnarray}
where $e_\nu(\vec{r}\,)$ is defined above by  Eq.~(\ref{enorm}).
In practice, we are usually interested only in the monopole term 
$Z(k)=Z(k;0,0)$, and will mostly use Eq.~(\ref{imp}) with replacement 
$e_\nu(\vec{s}\,) e_\nu(\vec{t}\:) \to \tilde{e}_\nu^2$, where 
$\tilde{e}_\nu \equiv e_\nu(0)$. In deriving Eq.~(\ref{imp}) we 
have neglected the coupling terms between $E_\nu$, $H_\tau$ and 
$H_z$, cf.\ Eqs.~(\ref{En})-(\ref{Hz}), which contribute to the 
third order of an expansion discussed below, and also have taken 
into account that $\psi_{\tau z}=\psi_{z \tau }$. 

For a small discontinuity, polarizabilities $\psi, \chi = O (h^3)$, 
and they are small compared to $b^3$. If we expand the impedance 
(\ref{imp}) in a perturbation series in polarizabilities, the first
order gives 
\begin{equation}
Z_1(k) = -\frac{ikZ_0 {\tilde{e}_\nu}^2 } {2}
 \left ( \psi_{\tau\tau}  -  \chi \right ) \ ,          \label{Z1}
\end{equation}
that is exactly the inductive impedance obtained in \cite{SK92p} for
an arbitrary cross section of the chamber. For a particular case 
of a circular pipe,  from either direct summation in 
(\ref{beamf}) or applying the Gauss law, one gets $\tilde{e}_\nu = 
1/(2\pi b)$. Substituting that into Eq.~(\ref{Z1}) leads to 
a well-known result \cite{SK92,RLG}:
\begin{equation}
Z(k) = - i k Z_0 \frac{\psi_{\tau\tau} - \chi}{8 \pi^2 b^2}
= - ikZ_0 \frac{\alpha_e+\alpha_m}{4\pi^2b^2} \ ,    \label{Zcirc}
\end{equation}
where we recall that two definitions of the polarizabilities are 
related as $\alpha_e=-\chi/2$ and $\alpha_m=\psi_{\tau\tau}/2$.

 From a physical point of view, keeping only the first order term 
(\ref{Z1}) corresponds to dropping out all radiation corrections 
in Eqs.~(\ref{En})-(\ref{Hz}).
These corrections first reveal themselves in the second order term
\begin{eqnarray}
Z_2(k)  = -\frac{ikZ_0 {\tilde{e}_\nu}^2 } {2} \left [  
 \psi^2_{\tau\tau} ( \Sigma_2 - \Sigma'_2 ) + 
 \psi^2_{\tau z} \Sigma'_3   \right.     \\   \label{Z2}
 \left. + \ \chi^2 (\Sigma'_1 - \Sigma_1) \ \right ] \nonumber \ ,        
\end{eqnarray}
which at frequencies above the chamber cutoff has both a real 
and imaginary part. The real part of the impedance is 
\begin{eqnarray}
Re Z_2(k)  & = & \frac{k^3Z_0 {\tilde{e}_\nu}^2 } {8}
 \left \{  \psi^2_{\tau z} \sum^{<}_s \frac{k'^2_s 
 \left (h^h_s \right)^2}{k^2 \beta'_s  }  \right. \label{ReZ} \\
 & + & \left. \psi^2_{\tau\tau}  \left [ \sum^{<}_s 
   \frac{\left (\nabla_\nu e^h_s \right)^2} { \beta_s k^2_s } 
 + \sum^{<}_s \frac{\beta'_s \left (\nabla_\tau h^h_s \right)^2} 
       {k^2 k'^2_s}  \right ] \right.    \nonumber  \\
 & + & \left. \chi^2 \left [ \sum^{<}_s \frac{\beta_s 
     \left (\nabla_\nu e^h_s \right)^2} {k^2 k^2_s}
 +  \sum^{<}_s \frac{ \left (\nabla_\tau h^h_s \right)^2} 
       {\beta'_s k'^2_s} \right ] \right \}  \ , \nonumber  
\end{eqnarray}
where the sums include only a finite number of the eigenmodes 
propagating in the chamber at a given frequency, i.e.\ those 
with $k_s<k$ or $k'_s<k$. 

The dependence of $Re\,Z$ on frequency is rather complicated; it 
has sharp peaks near the cutoffs of all propagating eigenmodes of 
the chamber, and increases in average with the frequency increase.
Well above the chamber cutoff, i.e.\ when $kb \gg 1$ (but still 
$kh \ll 1$ to justify the Bethe approach), this dependence can 
be derived as follows. If the waveguide cross section $S$ is
a simply connected region, the average number $n(k)$ of the 
eigenvalues $k_s$ (or $k'_s$) which are less than $k$, for 
$kb \gg 1$, is proportional to $k^2$ \cite{M&F}:
$$ n(k) \simeq \frac{S}{4 \pi} k^2 + O(k) \ , $$ where 
$S$ is the area of the cross section. Using this property, and 
taking into account that $\nabla_\nu e^h_s \propto k_s e^h_s$, and 
$\nabla_\tau h^h_s \propto k'_s h^h_s$, we replace sums in the RHS 
of Eq.~(\ref{ReZ}) by integrals as $\sum^{<}_s \to \int^k dk \frac
 {d}{dk}n(k)$. It turns out that all sums in Eq.~(\ref{ReZ}) have
the same asymptotic behavior, being linear in $k$, and as a result,
$Re\,Z \propto k^4$. Obtaining the exact coefficient in this 
dependence seems rather involved for a general $S$, but it can 
be easily done for a rectangular chamber, see in Appendix~B. 
The result is  
\begin{equation}
 Re\,Z = \frac{Z_0 k^4 {\tilde{e}_\nu}^2}{12 \pi }
 (\psi^2_{\tau\tau} + \psi^2_{\tau z} + \chi^2) \ . \label{ReZhf}
\end{equation}
Remarkably, the same answer (for $\psi_{\tau z}=0$) has been obtained 
in Ref.~\cite{SK92} simply by calculating the energy radiated by 
the dipoles into a half-space. The physical reason for this 
coincidence is clear: at frequencies well above the cutoff the 
effective dipoles radiate into the waveguide the same energy as 
into an open half-space.

Strictly speaking, the real part of impedance is non-zero even 
below the chamber cutoff, due to radiation outside.
In the case of a thin wall, $Re\,Z$ below the cutoff can be 
estimated by Eq.~(\ref{ReZhf}), and twice that for high frequencies, 
$kb \gg 1$. For a thick wall, the contribution of the radiation
outside to $Re\,Z$ is still given by Eq.~(\ref{ReZhf}), but with 
the outside polarizabilities substituted, and it decreases 
exponentially with the thickness increase \cite{RLG}.

The real part of the impedance is related to the power $P$
scattered by the hole into the beam pipe as $Re\,Z=2P/q^2$. 
These energy considerations can be used as an alternative way for 
the impedance calculation. The radiated power is
$$ P = \sum_s \left[ A^2_sP^{(E)}_s+B^2_sP^{(H)}_s \right] \ , $$ 
where we sum over all propagating modes in both directions, and 
$P_s$ means the time-averaged power radiated in $s$th eigenmode: 
$$P^{(E)}_s=k\beta_sk^2_s/(2Z_0) \quad \mbox{ and } \quad
P^{(H)}_s=Z_0k\beta'_sk'^2_s/2 \ . $$
Substituting beam fields (\ref{beamf}) into 
Eqs.~(\ref{Apm})-(\ref{cdq}) for the coefficients $A_s$ and $B_s$
and performing calculations gives us exactly the result (\ref{ReZ}). 
Such an alternative derivation of the real part has been 
carried out in Ref.~\cite{GS} for a circular pipe with a symmetric 
untilted hole ($\psi_{\tau z}=0$). Our result (\ref{ReZ}) coincides, 
in this particular case, with that of Reference \cite{GS}. 
It is appropriate to mention also that in this case at high 
frequencies the series has been summed approximately \cite{GS} 
using asymptotic expressions for roots of the Bessel functions, 
and the result, of course, agrees with Eq.~(\ref{ReZhf}).

One should note that the additional $\psi^2_{\tau z}$-term in 
Eq.~(\ref{ReZ}) is important in some particular cases. For 
example, this skew term gives a leading contribution to $Re\,Z$  
for a long and slightly tilted slot, because $\psi_{\tau z}$ 
can be much larger than $\psi_{\tau \tau}$ in this case, 
since $\psi_\| \gg \psi_\bot$, cf.\ Eqs.~(\ref{mpol}).

\subsection{Transverse Impedance}

We will make use of the expression for the generalized longitudinal 
impedance $Z(k;\vec s,\vec t \:)$, Eq.~(\ref{imp}). According to 
the Panofsky-Wenzel theorem, the transverse impedance can be derived 
as $\vec Z_\bot(k;\vec s,\vec t \:)= \vec {\nabla} 
Z(k;\vec s,\vec t\:)/(k s)$, see, e.g., \cite{SKrev} for details. 
This way leads to the expression
\begin{eqnarray}
&& \vec Z_\bot(k ; \vec s, \vec t \:)  =  - \frac{iZ_0 e^{dip}_{\nu}
 (\vec s\,) \vec {\nabla } e_{\nu}(\vec t\:)}{2 s} \times 
 \label{Ztgen} \\
&& \times \ \left [ \ \frac{\psi_{\tau\tau}}
 {1 - \psi_{\tau\tau} ( \Sigma_2 - \Sigma'_2 )} + 
 \psi^2_{\tau z} \Sigma'_3 - \frac{\chi}
 {1 - \chi ( \Sigma_1 - \Sigma'_1 )} \ \right ] \ , \nonumber
\end{eqnarray}
where $e^{dip}_{\nu}(\vec s\,) =
\vec s \cdot \vec {\nabla }e_{\nu}(\vec s\,)$. 

Going to the limit $s \to t \to 0$, we get the usual dipole 
transverse impedance
\begin{eqnarray}
&& \vec Z_\bot(k) = -iZ_0 \frac{1}{2}(d^2_x + d^2_y)\ \vec{a}_d 
 \cos (\varphi _b - \varphi _d)  \times                 \label{Zt} \\
&& \times \ \left [ \ \frac{\psi_{\tau\tau}}{1 - \psi_{\tau\tau} 
 ( \Sigma_2 - \Sigma'_2 )}
  +  \psi^2_{\tau z} \Sigma'_3 - \frac{\chi}
  {1 - \chi ( \Sigma_1 - \Sigma'_1 )} \ \right ] \ . \nonumber
\end{eqnarray}
Here $x,y$ are the horizontal and vertical coordinates in the
chamber cross section; $d_x \equiv \partial _x e_{\nu}(0)$,
$d_y \equiv \partial _y e_{\nu}(0)$; $\varphi _b =
\varphi _s =\varphi _t$ is the azimuthal angle of the beam position
in the cross-section plane; $\vec a_d =
\vec a_x \cos \varphi _d + \vec a_y \sin \varphi _d$ is a unit vector
in this plane in direction $\varphi_d$, which is defined by 
conditions $\cos \varphi _d = d_x/\sqrt{\,d^2_x + d^2_y}$,
$\sin \varphi _d = d_y/\sqrt{\,d^2_x + d^2_y}$. It is seen from 
Eq.~(\ref{Zt}) that the angle $\varphi _d$ shows the direction of the
transverse-impedance vector $\vec Z_\bot$ and, therefore, of the
beam-deflecting force. Moreover, the value of $Z_\bot$ is maximal
when the beam is deflected along this direction and vanishes when
the beam offset is perpendicular to it. 

The equation (\ref{Zt}) includes the corrections due to waves 
radiated by the hole into the chamber in exactly the same way as 
Eq.~(\ref{imp}) for the longitudinal impedance. If we expand it
in a series in the polarizabilities, the first order of the square 
brackets in (\ref{Zt}) gives $(\psi_{\tau\tau} -\chi)$, and the 
resulting inductive impedance is \cite{SK92p}:
\begin{equation}
 \vec Z_\bot(k) = -iZ_0 \frac{1}{2}(d^2_x + d^2_y)\ \vec{a}_d 
 \cos (\varphi _b - \varphi _d)     
\left ( \psi_{\tau\tau} - \chi \right ) \ . \label{Zt1}
\end{equation}

For a circular pipe, $d_x=\cos \varphi_h/(\pi b^2)$ and 
$d_y=\sin \varphi_h/(\pi b^2)$,
where $\varphi_h$ is the azimuthal position of the discontinuity 
(hole). As a result, $\varphi _d = \varphi _h$, and the 
deflecting force is directed toward (or opposite to) the hole. 
Note that in axisymmetric structures the beam-deflecting force 
is directed along the beam offset; the presence of an obstacle 
obviously breaks this symmetry.
For a general cross section, the direction of the deflecting 
force depends on the hole position in a complicated way,
see \cite{SK92p} for rectangular and elliptic chambers. 

The transverse impedance of a discontinuity
on the wall of a circular pipe has a simple form \cite{SK92}:
\begin{equation}
 \vec Z_{\bot\; circ}(k) = -iZ_0 \frac{\psi_{\tau\tau} - \chi}
 {2\pi^2 b^4}\ \vec{a}_h \cos (\varphi _b - \varphi _d) =    
 -iZ_0 \frac{\alpha_m + \alpha_e} {\pi^2 b^4}\ 
 \vec{a}_h \cos (\varphi _b - \varphi _d) \ ,      \label{Zt1circ}
\end{equation}
where $\vec{a}_h$ is a unit vector in the direction from the 
chamber axis to the discontinuity, orthogonal to $\hat{z}$.

For $M$ ($M\ge 3$) holes uniformly spaced in one cross
section of a circular beam pipe, a vector sum of M expressions 
(\ref{Zt1circ}) gives the transverse impedance as
\begin{equation}
\vec Z_{\bot\; circ\; M}(k) = -iZ_0 \frac{\alpha _m + \alpha _e}
{\pi ^2b^4} \frac{M}{2} \vec{a}_b \ ,
\end{equation}
where $\vec{a}_b=\vec s/|\vec s|$ is a unit vector in the direction 
of the beam transverse offset. One can see that the deflecting force 
is now directed along the beam offset, i.e.\ some kind of the axial
symmetry restoration occurs. The maximal value of $Z_\bot$ for
$M$ holes which are uniformly spaced in one cross-section is only
$M/2$ times larger than that for $M=1$. Moreover, the well-known
empirical relation $Z_\bot = (2/b^2k)Z$, which is justified only 
for axisymmetric structures, holds in this case also.

 The second order term in Eq.~(\ref{Zt}) includes $Re\,Z_\bot$, 
cf.\ Sect.~3.1 for the longitudinal impedance.

\section{Trapped Modes}

So far we considered the perturbation expansion of Eq.~(\ref{imp})
implicitly assuming that correction terms $O(\psi)$ and $O(\chi)$ 
in the denominators of its right-hand side (RHS) are small compared 
to 1. Under certain conditions this assumption is incorrect, 
and it leads to some non-perturbative results. Indeed, 
at frequencies slightly below the chamber cut-offs, $0 <  k_s - k 
 \ll k_s$ (or the same with replacement $k_s \to k'_s$), a 
single term in sums $\Sigma'_1$, $\Sigma_2$, or $\Sigma'_3$ becomes
very large, due to very small $\Gamma_s=(k_s^2-k^2)^{1/2}$ (or 
$\Gamma'_s$) in its denominator, and then the ``corrections'' 
$\psi \Sigma$ or $\chi \Sigma$ can be of the order of 1. As a result,
one of the denominators of the RHS of Eqs.~(\ref{imp}) can vanish, 
which corresponds to a resonance of the coupling impedance. 
On the other hand, vanishing denominators in 
Eqs.~(\ref{En})-(\ref{Hz}) mean the existence of 
non-perturbative eigenmodes of the chamber with a hole, since 
non-trivial solutions $E,H \neq 0$ exist even for vanishing external
(beam) fields $E^b,H^b = 0$. These eigenmodes are nothing but
the trapped modes studied in \cite{S&K94} for a circular waveguide 
with a small discontinuity. In our approach, one can easily derive
parameters of trapped modes for waveguides with an arbitrary cross 
section. 

\subsection{Frequency Shifts}

Let us for brevity restrict ourselves to the case $\psi_{\tau z}=0$ 
and consider Eq.~(\ref{Ht}) in more detail. For $H^b = 0$ we have  
\begin{equation}
H_\tau \left [ 1 - \psi_{\tau\tau} \frac { k^2 \left 
 (\nabla_\nu e^h_s \right)^2} {4 \Gamma_s k^2_s } 
 + \ldots \right ] = 0 \ ,             \label{Htm}
\end{equation}
where $s \equiv \{nm\}$ is the generalized index, and $\ldots$ mean 
all other terms of the series $\Sigma_2,\Sigma'_2$. At frequency 
$\Omega_s $ slightly below the cutoff frequency $\omega_s= k_s c$ 
of the TM${}_s$-mode, the fraction in Eq.~(\ref{Htm}) is large due 
to small $\Gamma_s$ in its denominator, and one can neglect the 
other terms. Then the condition for a non-trivial solution 
$H_\tau \neq 0$ to exist is
\begin{equation}
\Gamma_s \simeq \frac {1}{4} \psi_{\tau\tau} \left 
 (\nabla_\nu e^h_s \right)^2 \ .                \label{GamE}
\end{equation}
In other words, there is a solution of the homogeneous, 
i.e., without external currents, Maxwell equations for the chamber 
with the hole, having the frequency $\Omega_s < \omega_s$ --- the 
$s$th trapped TM-mode. When Eq.~(\ref{GamE}) is satisfied, the 
series (\ref{fexp}) is obviously dominated by the single term 
$A_sF^{E}_s$; hence, the fields of the trapped mode have the form
[cf. Eq.~(\ref{emode})]
\begin{eqnarray}
{\cal{E}}_z & = & k_s^2 e_s \exp(-\Gamma _s|z|) \ ; 
     \qquad  {\cal{H}}_z = 0 \ ; \nonumber \\
\vec{\cal{E}}_t & = & \mbox{sgn}(z) \Gamma _s \vec{\nabla}e_s
        \exp(- \Gamma _s|z|) \ ;               \label{temode} \\
Z_0 \vec{\cal{H}}_t & = & ik \hat{z} \times \vec{\nabla} 
   e_s \exp(- \Gamma _s|z|) \ , \nonumber 
\end{eqnarray}
up to some arbitrary amplitude. Strictly speaking, 
these expressions are valid at distances $|z| > b$ from 
the discontinuity. Typically, $\psi_{\tau\tau}= O(h^3)$ and  
$\nabla_\nu e^h_s = O(1/b)$, and, as a result, $\Gamma_sb \ll 1$. 
It follows that the field of the trapped mode extends along the 
vacuum chamber over the distance $1/\Gamma_s$, large compared
to the chamber transverse dimension $b$. 

The existence of the trapped modes in a circular wave\-guide with 
a small hole was first proved in \cite{S&K94}, and conditions similar 
to Eq.~(\ref{GamE}) for this particular case were obtained in 
\cite{S&K94,K95}, using the Lorentz reciprocity theorem. From 
the general approach presented here for the wave\-guide with an 
arbitrary cross section, their existence follows in a natural way. 
Moreover, in such a derivation, the physical mechanism of this 
phenomenon becomes quite clear: a tangential magnetic field induces 
a magnetic moment on the hole, and the induced magnetic moment 
supports this field if the resonance condition (\ref{GamE}) is 
satisfied, so that the mode can exist even without an external 
source. One should also note that the induced electric moment 
$P_\nu$ is negligible for the trapped TM-mode, since 
$P_\nu = O(\Gamma_sb)M_\tau$, as follows from Eq.~(\ref{temode}). 

The equation (\ref{GamE}) gives the frequency shift $\Delta \omega_s 
\equiv \omega_s  - \Omega_s$ of the trapped $s$th TM-mode down from 
the cutoff $\omega_s$
\begin{equation}
 \frac{\Delta \omega_s}{\omega_s} \simeq \frac{1}{32 k^2_s} 
  \psi^2_{\tau\tau} \left (\nabla_\nu e^h_s \right)^4 \ .  \label{dwE}
\end{equation}
In the case of a small hole this frequency shift is very small, and 
for the trapped mode (\ref{temode}) to exist, the width of the 
resonance should be smaller than $\Delta \omega_s $. Contributions to 
the resonance width come from energy dissipation in the waveguide wall 
due to its finite conductivity, and from energy radiation inside the 
waveguide and outside, through the hole. Radiation escaping through 
the hole is easy to estimate \cite{S&K94}, and for a thick wall it is 
exponentially small, e.g., \cite{RLG}. The damping rate due to a finite 
conductivity is $\gamma = P/(2W)$, where $P$ is the time-averaged power 
dissipation and $W$ is the total field energy in the trapped mode, 
which yields 
\begin{equation}
 \frac{\gamma_s}{\omega_s} = \frac{\delta}{4 k^2_s} 
 \oint dl \left (\nabla_\nu e_s \right )^2 \ ,  \label{gamE}
\end{equation}
where $\delta$ is the skin-depth at frequency $\Omega_s$, 
and the integration is along the boundary ${\partial S}$.
The evaluation of the radiation into the lower waveguide modes 
propagating in the chamber at given frequency $\Omega_s$ is also 
straightforward \cite{GS}, if one makes use of the coefficients of 
mode excitation by effective dipoles on the hole, 
Eqs.~(\ref{Apm})-(\ref{cdq}). The corresponding damping rate
$\gamma_R = O(\psi^3)$ is small compared to $\Delta \omega_s $. 
For instance, if there is only one TE${}_p$-mode with the frequency  
below that for the lowest TM${}_s$-mode, like in a circular waveguide 
(H${}_{11}$ has a lower cutoff than E${}_{01}$), 
\begin{equation}
 \frac{\gamma_R}{\Delta \omega_s} = \frac{\psi_{\tau \tau} \beta'_p}
 {k'^2_p} \left (\nabla_\nu h^h_s \right )^2 \ ,  \label{gamER}
\end{equation}
where $\beta'_p \simeq (k_s^2-k'^2_p)^{1/2}$ because $k \simeq k_s$.

One can easily see that denominator $[1 - \chi ( \Sigma_1 - 
\Sigma'_1 )]$ in Eq.~(\ref{En}) does not vanish because singular
terms in $\Sigma'_1$ have a ``wrong'' sign. However, due to the 
coupling between $E_\nu$ and $H_z$, a non-trivial solution 
$E_\nu,H_z \neq 0$ of simultaneous equations (\ref{En}) 
and (\ref{Hz}) can exist, even when $E^b = 0$. The corresponding 
condition has the form 
\begin{equation}
\Gamma'_{nm} \simeq \frac{1}{4} \left [ \psi_{zz} k'^2_{nm} 
 \left ( h^h_{nm} \right)^2 - \chi \left 
 (\nabla_\tau h^h_{nm} \right)^2 \right ] \ ,      \label{GamH}
\end{equation}
which gives the frequency of the trapped TE${}_{nm}$-mode, 
provided the RHS of Eq.~(\ref{GamH}) is positive.

\subsection{Impedance}

The trapped mode (\ref{temode}) gives a resonance contribution to 
the longitudinal coupling impedance at $\omega \approx \Omega_s$
\begin{equation}
 Z_s(\omega ) = \frac{2i \Omega_s \gamma_s R_s}
                     {\omega^2 - (\Omega_s -i\gamma_s)^2} \ ,
                   \label{trimp}
\end{equation}
where the shunt impedance $R_s$ can be calculated as that for a 
cavity with given eigenmodes, e.g.~\cite{SKrev},
\begin{equation}
R_s = {\sigma \delta  \left| \int{dz \exp (-i \Omega_s z /c) 
{\cal E}_{z}(z)} \right|^2 \over \int_{S_w} ds 
|{\cal H}_{\tau }|^2 } \ .
                 \label{Rimp}
\end{equation}
The integral in the denominator is taken over the inner wall surface, 
and we assume here that the power losses due to its finite 
conductivity dominate. Integrating in the numerator one should 
include all TM-modes generated by the effective magnetic moment 
on the hole using Eqs.~(\ref{Apm})-(\ref{cdq}), in spite of a 
large amplitude of only the trapped TM${}_s$ mode. While all other 
amplitudes are suppressed by factor $\Gamma_sb \ll 1$, their 
contributions are comparable to that from TM${}_s$, because
this integration produces the factor $\Gamma_q b$ for any 
TM${}_q$ mode. The integral in the denominator is obviously 
dominated by TM${}_s$. Performing calculations yields 
\begin{equation}
R_s = \frac{Z_0 {\tilde{e}}^2_\nu \psi^3_{\tau\tau} k_s
 \left (\nabla_\nu e^h_s \right)^4} {8 \delta 
 \oint dl \left (\nabla_\nu e_s \right )^2  } \ ,   \label{Rs}
\end{equation}
where $\tilde{e}_\nu = e_\nu(0)$ is defined by Eq.~(\ref{enorm}).

Results for a particular shape of the chamber cross section 
can be obtained from the equations above by substituting the 
corresponding eigenfunctions (see Appendix).

One should note that typically the peak value $R_s$ of the impedance 
resonance due to one small hole is rather small except for the 
limit of a perfectly conducting wall, $\delta \to 0$ --- indeed, 
$R_s \propto (h/b)^9b/\delta$, and $h \ll b$. However, for many 
not-so-far separated holes, the resulting impedance can be much 
larger. The trapped modes for many discontinuities on a circular 
waveguide has been studied in Ref.~\cite{K95}, and the results can 
be readily transferred to the considered case of an arbitrary shape 
of the chamber cross section. In particular, it was demonstrated 
that the resonance impedance in the extreme case can be as large 
as $N^3$ times that for a single discontinuity, where $N$ is 
the number of discontinuities. It strongly depends on the 
distribution of discontinuities, or on the distance between them 
if a regular array is considered. 

After the trapped modes in beam pipes with small holes were predicted 
theoretically \cite{S&K94,K95}, their existence was proved by 
experiments with perforated waveguides at CERN \cite{C&S95}.

\section{Analytical Formulae for Some Small Discontinuities} 

For reader convenience, in this section we collected analytical 
expressions for the coupling impedances of various small 
discontinuities. The expressions give the inductive part of
the impedance and work well at frequencies below the chamber 
cutoff, and, in many cases, even at much higher frequencies.
However, there can also exist resonances of the real part at 
frequencies near the cutoff for holes and cavities due to the 
trapped modes, as was shown in \cite{S&K94,K95}, and the real part 
of the impedance due to the energy radiated into the beam pipe should 
be taken into account at frequencies above the cutoff, 
see \cite{SK95,KGS,K&YHC}.

It is worth noting that both the longitudinal and 
transverse impedance are proportional to the same combination of 
polarizabilities, $\alpha_e + \alpha_m$, for any cross section of
the beam pipe. Here we use the effective polarizabilities $\alpha_e, 
\alpha_m$ as defined in \cite{Collin}; they are related to the 
magnetic susceptibility $\psi$ and the electric polarizability 
$\chi$ of an obstacle as $\alpha_e=-\chi/2$ and $\alpha_m=\psi/2$. 
The real part of the impedance is proportional to 
$\alpha^2_e + \alpha^2_m$, and is usually small compared to 
the reactance at frequencies below the chamber cutoff. 
While the impedances below are written for a round pipe, more 
results for the other chamber cross sections, the impedance 
dependence on the obstacle position on the wall and on the beam 
position can be found in \cite{SK92p,SK95,KGS,K&YHC,SK96e}.

The longitudinal impedance of a small obstacle on the wall of a 
cylindrical beam pipe with a circular transverse cross section of 
radius $R$ is simply \cite{SK92} 
(up to notations, it is the same Eq.~(\ref{Zcirc}) above)
\begin{equation}
Z(k) = - i k Z_0 \frac{\alpha_e + \alpha_m}{4 \pi^2 R^2} \ , \label{Z}
\end{equation}
where $Z_0 = 120 \pi$~Ohms is the impedance of free space, 
$k=\omega/c$ is the wave number, and $\alpha_e, \alpha_m$ are the 
electric and magnetic polarizabilities of the discontinuity. 
The polarizabilities depend on the obstacle shape and size.

The transverse dipole impedance of the discontinuity for this case is 
\begin{equation}
\vec Z_\bot(\omega) = -iZ_0 \frac{\alpha _m + \alpha _e}
{\pi ^2 R^4} \vec a_h \cos (\varphi _h-\varphi _b) \ ,  \label{Ztcm}
\end{equation}
where $\vec a_h$ is the unit vector directed to the obstacle in 
the chamber transverse cross section containing it,  
$\varphi _h$ and $\varphi _b$ are azimuthal angles of the 
obstacle and beam in this cross section.

\subsection{Holes and Slots}

For a circular hole with radius $a$ in a thin wall, when 
thickness $t \ll a$, the polarizabilities are 
$\alpha _{m} = 4a^3/3, \ \alpha _e =  - 2a^3/3$, so that
the impedance Eq.~(\ref{Z}) takes a simple form
\begin{equation}
Z(k) = - i k Z_0 \frac{a^3}{6 \pi^2 R^2} \ , \label{Zh}
\end{equation}
and similarly for Eq.~(\ref{Ztcm}). 
For the hole in a thick wall, $t \ge a$, the sum 
$(\alpha _m + \alpha _e)=2a^3/3$ should be
multiplied by a factor 0.56, see \cite{RLG,G&D}. 
There are also analytical expressions for polarizabilities 
of elliptic holes in a thin wall \cite{Collin}, and 
paper \cite{R&G} gives thickness corrections for 
this case. Surprisingly, the thickness factor for 
$(\alpha _m + \alpha _e)$ exhibits only a weak dependence 
on ellipse eccentricity $\varepsilon$, changing its 
limiting value for the thick wall from 0.56 for $\varepsilon=0$ 
to 0.59 for $\varepsilon=0.99$. 

For a longitudinal slot of length $l$ and width $w$, $w/l \le 1$, 
in a thin wall, useful approximations have been obtained 
\cite{SK93s}: for a rectangular slot 
$$\alpha _m + \alpha _e = w^3 (0.1814 - 0.0344 w/l) \ ;$$
and for a rounded end slot 
$$\alpha _m + \alpha _e = w^3 (0.1334 - 0.0500 w/l) \ ; $$
substituting of which into Eqs.~(\ref{Z})-(\ref{Ztcm}) gives
the impedances of slots. Figure~1 compares impedances
for different shapes of pumping holes. 

\begin{figure}
\centerline{\epsfig{figure=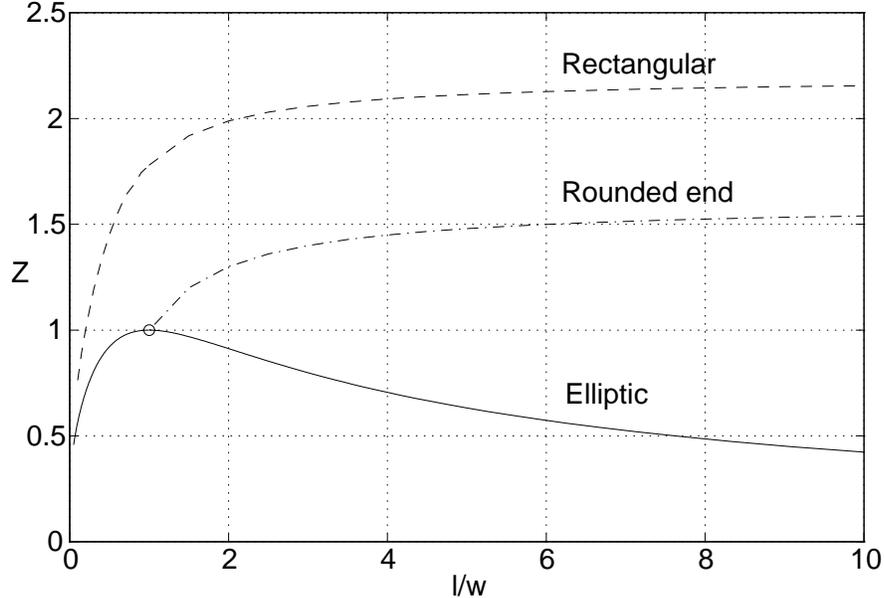,width=12cm}}
\caption{Slot impedance versus slot length $l$ for fixed  
width $w$ in units of the impedance of the circular hole with
diameter $w$.}
\end{figure}

\subsection{Annular Cut}

The polarizabilities of a ring-shaped cut in the wall of an 
arbitrary thickness have been calculated in \cite{SK96a}. 
Such an aperture can serve as an approximation for a electrode 
of a button-type beam position monitor, for a thin wall, or 
a model of a coax attached to the vacuum chamber, when the 
wall thickness is large. If $a$ 
and $b$ denote the inner and outer radii of the annular cut, 
$a \le b \ll R$, the magnetic susceptibility of a 
narrow ($w = b-a \ll b$) annular slot in a thin plate is 
\begin{equation}
 \psi \simeq \frac {\pi^2 b^2 a } {\ln (32b/w) -2} \ . \label{psin}
\end{equation}
For a narrow annular gap in the thick wall the asymptotic 
behavior is $\psi \simeq 2 \pi b^2 w $. 
The approximation (\ref{psin}) works well for narrow gaps, 
$w/b \le 0.15$, while the thick wall result is good only for
$w/b \le 0.05$.

Analytical results for the electric polarizability 
of a narrow annular cut are:  for a thin wall 
\begin{equation}
\chi \simeq \pi^2 w^2 (b+a)/8 \ ,  \label{chianal0}
\end{equation}
and for a thick wall
\begin{equation}
\chi \simeq w^2 (b+a) \ .  \label{chianalt}
\end{equation}
These estimates work amazingly well even for very wide gaps, 
up to $w/b \ge 0.85$. The electric polarizability depends 
on the wall thickness rather weakly. 

The difference $(\psi - \chi)/b^3=2(\alpha _m + \alpha _e)/b^3$ 
for an annular cut, calculated by variational methods in 
\cite{SK96a}, is plotted in Fig.~2 for a few values of the wall 
thickness. 

\begin{figure}
\centerline{\epsfig{figure=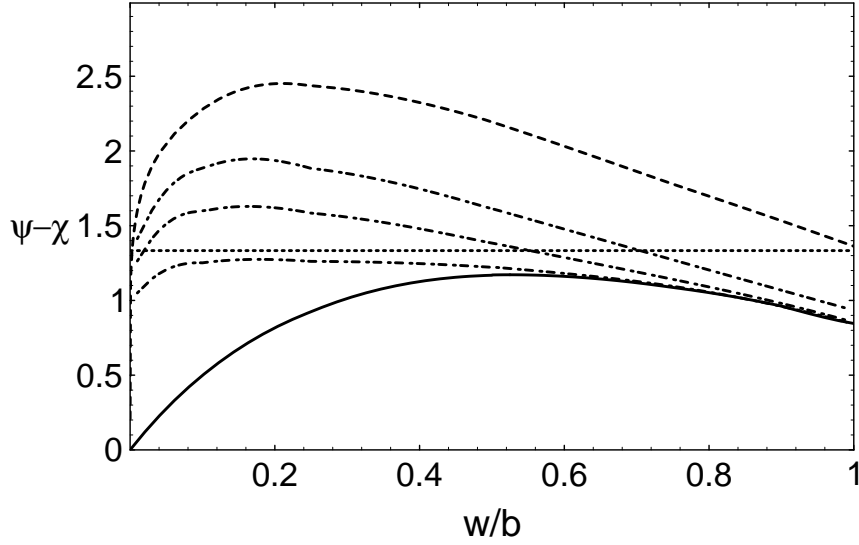,width=12cm}}
\caption{Difference of the polarizabilities (in units of 
$b^3$) of an annular cut versus its relative width $w/b$ for 
different thicknesses of the wall $t=0;\, w/2;\, w;\, 2w$, and
$t \gg w$ (from top to bottom). The dotted line corresponds to 
the circular hole in a thin wall, $(\psi - \chi)/b^3=4/3$.}
\end{figure}

\subsection{Protrusions}

For a protrusion inside the beam pipe having the 
shape of a half ellipsoid with semiaxis $a$ in the longitudinal 
direction (along the chamber axis), $b$ in the radial direction, 
and $c$ in the azimuthal one, with $a, b, c \ll R$, 
the polarizabilities are \cite{SK96}
\begin{equation}
\alpha_e = 
 \frac{2 \pi a b c}{3 I_b} \ ,            \label{ae}
\end{equation}
and 
\begin{equation}
\alpha_m = 
 \frac{2 \pi a b c}{3 (I_c - 1)} \ ,       \label{am}
\end{equation}
where
\begin{equation}
I_b = \frac{abc}{2} \int_0^\infty \frac{ds}
 { (s+b^2)^{3/2}(s+a^2)^{1/2}(s+c^2)^{1/2} } \ ,  \label{I}
\end{equation}
and $I_c$ is given by Eq.~(\ref{I}) with $b$ and $c$ interchanged.

In the particular case $a=c$, $b=h$ we have an ellipsoid of 
revolution, and the polarizabilities are expressed in 
terms of the hypergeometric function ${}_2F_1$:
\begin{equation}
\alpha_e =  \frac{2 \pi a^2 h}
              {{}_2F_1(1,1;5/2;1-h^2/a^2)} \ ,      \label{ae1}
\end{equation}
and 
\begin{equation}
\alpha_m =  \frac{2 \pi a^2 h}
 {{}_2F_1(1,1;5/2;1-a^2/h^2) - 3 } \ . \label{am1}
\end{equation}

\subsubsection{Post} 

In the limit $a=c \ll h$, corresponding to a pinlike obstacle, we get 
a simple expression for the 
inductive impedance of a narrow pin (post) of height $h$ and radius 
$a$, protruding radially into the beam pipe:\footnote{One 
could use the known result for the induced electric dipole of a narrow
cylinder parallel to the electric field \cite{Landau}. It will only
change $\ln(2h/a) - 1$ in Eq.~(\ref{Zpin}) to $\ln(4h/a) - 7/3$.}
\begin{equation}
Z(k) \simeq - i k Z_0 \frac{h^3}
     {6 \pi R^2 \left(\ln\,(2h/a) - 1 \right) } \ .    \label{Zpin}
\end{equation}
The factor $F \equiv (\alpha_e + \alpha_m)/V$, where $V=2\pi a^2 h/3$ 
is the volume occupied by the obstacle, is plotted in Fig.~3 
versus the ratio $h/a$. The figure also shows comparison with the 
asymptotic approximation given by Eq.~(\ref{Zpin}).

\begin{figure}
\centerline{\epsfig{figure=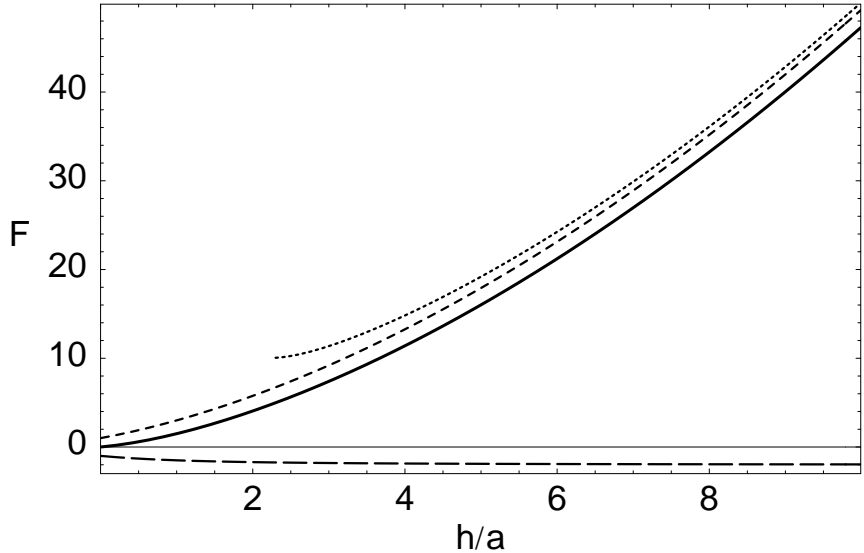,width=12cm}}
\caption{Function $F \equiv (\alpha_e + \alpha_m)/V$ versus 
aspect ratio $h/a$ for a pinlike obstacle (solid line). 
The electric contribution is short dashed, the magnetic one is 
long dashed, and the dotted line shows the asymptotic form
used in Eq.~(\ref{Zpin}).}
\end{figure}
 
\subsubsection{Mask} 

One more particular case of interest is $h=a$, i.e. a 
semispherical obstacle of radius $a$. From 
Eqs.~(\ref{ae1})-(\ref{am1}) the impedance of such a discontinuity is
\begin{equation}
Z(k) = - i k Z_0 \frac{a^3} {4 \pi R^2 } \ ,         \label{Zss}
\end{equation}
which is $3\pi/2$ times that for a circular hole of the same radius
in a thin wall, cf.\ Eq.~(\ref{Zh}).

Another useful result that can be derived from the general solution,
Eqs.~(\ref{ae})-(\ref{I}), is the impedance of a mask intended to
intercept synchrotron radiation. We put $b=c=h$, so that our model mask
has the semicircular shape with radius $h$ in its largest transverse 
cross section. Then the integral in Eq.~(\ref{I}) is reduced to 
$$I_b=I_c= \frac{1}{3} \, {}_2F_1 \left(1,\frac{1}{2};\frac{5}{2};
 1-\frac{h^2}{a^2}\right) \ ,$$ 
and we can further simplify the result for two particular cases. 

The first one is the thin mask, $a \ll h$,
in which case $\alpha_e \simeq  8h^3/3$ , and again it dominates the
magnetic term, $\alpha_m \simeq  -V = -2\pi a h^2/3$. The coupling 
impedance for such an obstacle 
--- a half disk of radius $h$ and thickness $2a$, $a \ll h$, 
transverse to the chamber axis --- is therefore
\begin{equation}
Z(k) = - i k Z_0 \frac{2 h^3} {3 \pi^2 R^2 } \left [ 1 + 
  \left( \frac{4}{\pi} - \frac{\pi}{4} \right) \frac{a}{h}
  + \ldots \, \right ] \, ,        \label{Zm0}
\end{equation}
where the next-to-leading term is shown explicitly. 

In the opposite limit, $h \ll a$, which corresponds to a long (along 
the beam) mask, the leading terms $\alpha_e \simeq - \alpha_m \simeq  
4\pi a h^2/3$ cancel each other. As a result, the impedance 
of a long mask with length $l=2a$ and height $h$,  $h \ll l$, is
\begin{equation}
Z(k) \simeq - i k Z_0 \frac{4 h^4}{3 \pi R^2 l } 
    \left(\ln \frac{l}{h} - 1 \right) \ ,               \label{Zml}
\end{equation}
which is relatively small due to the ``aerodynamic'' shape of this 
obstacle, in complete analogy with results for long elliptic slots 
\cite{SK92,RLG,SK93s}. 

Figure~4 shows the impedance of a mask with a given semicircular 
transverse cross section of radius $h$ versus its normalized 
half length, $a/h$. The comparison with the asymptotic approximations 
Eqs.~(\ref{Zm0}) and (\ref{Zml}) is also shown. One can see, that
the asymptotic behavior (\ref{Zml}) starts to work well only for very
long masks, namely, when $l = 2a \ge 10h$. Figure~4 demonstrates
that the mask impedance depends rather weakly on the length. 
Even a very thin mask ($a \ll h$) contributes as much as $8/(3\pi) 
 \simeq 0.85$ times the semisphere ($a=h$) impedance, Eq.~(\ref{Zss}),
while for long masks the impedance decreases slowly: at $l/h=20$, it
is still 0.54 of that for the semisphere. 

\begin{figure}
\centerline{\epsfig{figure=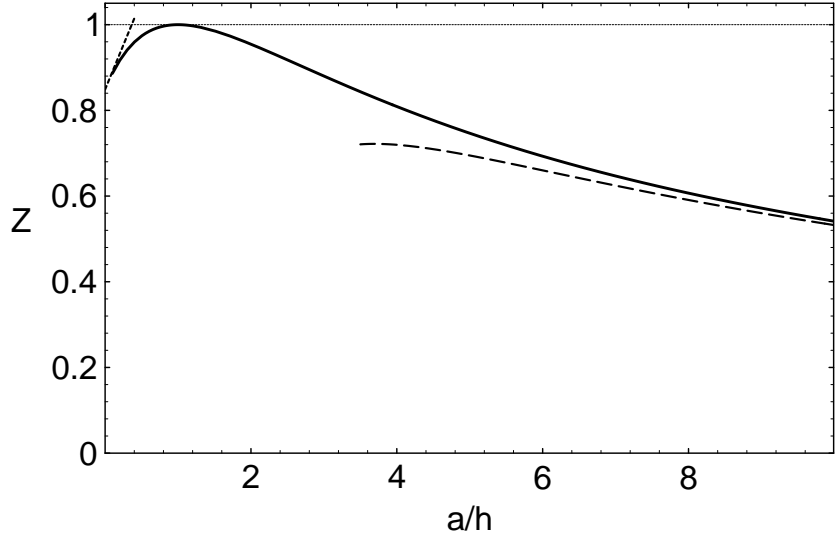,width=12cm}}
\caption{Impedance $Z$ of a mask (in units of that for a semisphere 
with the same depth, Eq.~(\ref{Zss}) with $a=h$) versus its length. 
The narrow-mask approximation, Eq.~(\ref{Zm0}), is short dashed, and 
the long-mask one, Eq.~(\ref{Zml}), is long dashed.}
\end{figure}
 
In practice, however, the mask has usually an abrupt cut toward the 
incident synchrotron radiation, so that it is rather one-half of a
long mask. From considerations above one can suggest as a reasonable 
impedance estimate for such a discontinuity the half sum of the 
impedances given by Eqs.~(\ref{Zm0}) and (\ref{Zml}). This estimate 
is corroborated by 3D numerical simulations using the MAFIA code, 
at least, for the masks which are not too long. 
In fact, the low-frequency impedances of a
semisphere and a half semisphere of the same depth --- which can 
be considered as a relatively short realistic mask --- were found 
numerically to be almost equal (within the errors), and close to
that for a longer half mask. From these results one can conclude 
that a good estimate for the mask impedance is given simply
by Eq.~(\ref{Zss}). 

\subsection{Axisymmetric Discontinuities} 

Following a similar procedure one can also easily obtain the results 
for axisymmetric irises having a semi-elliptic profile in the 
longitudinal chamber cross section, with depth $b=h$ and length 
$2a$ along the beam. 
For that purpose, one should consider limit $c \to \infty$ in
Eq.~(\ref{I}) to calculate the effective polarizabilities 
$\tilde{\alpha}_e$ and 
$\tilde{\alpha}_m$ per unit length of the circumference of the
chamber transverse cross section. The broad-band impedances
of axisymmetric discontinuities have been studied in \cite{K&S},
and the longitudinal coupling impedance is given by
\begin{equation}
Z(k) = - i k Z_0 \frac{\tilde{\alpha}_e + 
         \tilde{\alpha}_m}{2 \pi R} \ ,                \label{Zax}
\end{equation}
quite similar to Eq.~(\ref{Z}). As $c \to \infty$, the integral
$I_c \to 0$, and $I_b$ is expressed in elementary functions as
$$I_b=\frac{1}{2}{}_2F_1 \left(1,\frac{1}{2};2;
 1-\frac{h^2}{a^2}\right) =\frac{a}{a+h} \ . $$
It gives us immediately
\begin{equation}
 \tilde{\alpha}_e = \frac{\pi}{2}\, h(h+a) \, ; \qquad
  \tilde{\alpha}_m = -\frac{\pi}{2}\, ah \ ,         \label{alax}
\end{equation}
and the resulting impedance of the iris of depth $h$ with the 
semielliptic profile is simply
\begin{equation}
Z(k) = - i k Z_0 \frac{h^2} {4R } \ ,               \label{Ziris}
\end{equation}
which proves to be independent of the iris thickness $a$. The same
result has been obtained using another method \cite{RLGu},
and also directly by conformal mapping in \cite{SK96}, following
the general method of \cite{K&S}. 

Using a conformal mapping, one can readily obtain an answer also for 
irises  having the profile shaped as a circle segment with the chord 
of length $s$ along the chamber wall in the longitudinal direction, 
and opening angle $2\varphi$, where $0 \le \varphi \le \pi$. 
The impedance of such an exotic iris, expressed 
in terms of its height $h=s(1-\cos{\varphi})/(2\sin{\varphi})$: 
\begin{eqnarray}
Z(k) & = & - i k Z_0 \frac{h^2} 
       {2R (1-\cos{\varphi})^2 } \times             \label{Zsegm} \\
 & \times & \left [ \frac{\varphi (2\pi-\varphi)}{3(\pi-\varphi)^2} 
 \sin^2{\varphi}
  - \frac{2\varphi - \sin{2\varphi}}{2\pi} \right ] \ .  \nonumber
\end{eqnarray}
Again, the impedance is proportional to $h^2$, but the coefficient 
now depends (in fact, rather weakly) on $\varphi$.

A few useful results for low-frequency impedances of axisymmetric
cavities and irises with a rectangular, trapezoidal and triangular
transverse profile have been obtained in \cite{K&S} using conformal
mapping to calculate the electric polarizability.

The low-frequency impedance of the small short pill-box whose 
length $g$ is not large than depth $h$ is 
\begin{equation}
Z(\omega) = -i k Z_0 \frac{1}{2\pi R}
\left (gh-\frac{g^2}{2\pi}\right ) \ ,            \label{Zpb}
\end{equation}

The low-frequency impedance of the shallow enlargement 
takes the form 
\begin{equation}
Z(\omega) = -i k Z_0 \frac{h^2}{2\pi ^2 R}
\left ( 2\ln (2\pi g/h) +1\right ) \ ,           \label{Zshl}
\end{equation}
where $g \gg h$, but still less than $R$. 

The low-frequency impedance of a small step of depth $h \ll R$ is 
\begin{equation}
Z(\omega) = -i k Z_0 \frac{h^2}{4\pi ^2 R}
\left ( 2\ln (2\pi R/h) +1\right ) \ .            \label{Zstep}
\end{equation}

The inductance produced by the transition with the slope 
angle $\theta= \pi \nu$ has the form: 
\begin{equation}
Z(\omega) = -i \frac{Z_0 k h^2}{2\pi ^2 R}
\left \{ \ln \left [ \pi \nu \left (\frac{b}{h} - 
          2\cot \pi \nu \right ) \right ] +
\frac{3}{2} - \gamma -\psi (\nu) -\frac{\pi }{2}\cot \pi \nu - 
\frac{1}{2\nu}  \right \},                        \label{Zslop}
\end{equation}
where $\gamma = 0.5772 \ldots$ is Euler's constant, $\psi (\nu)$ is 
the psi-function and the transition is assumed to be short compared 
to the chamber radius, i.e.\  transition length 
$l=h\cot \pi \nu \ll R$. The ratio of this inductance to 
that of the abrupt step ($\nu=1/2$, Eq.~(\ref{Zstep})) with the same 
height is plotted in Fig.~5 as a function of the slope angle. 

\begin{figure}
\centerline{\epsfig{figure=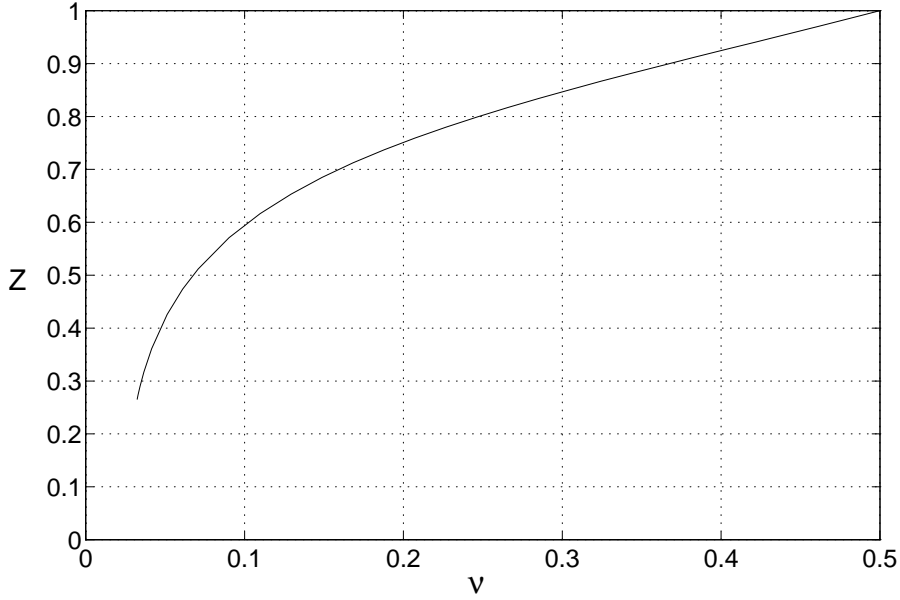,width=12cm}}
\caption{Impedance $Z$ of a transition versus its slop, in units
of that for the abrupt step ($\nu=1/2$, Eq.~(\ref{Zstep})) with 
the same height.} 
\end{figure}

The impedance of a thin (or deep) iris, $g\ll h$, has the form
\begin{equation}
Z(\omega) = -i k Z_0 \frac{1}{4 R} \left [ h^2 + 
\frac{gh}{\pi} \left ( \ln (8\pi g/h) - 3 \right ) \right ] \ .         
                                                      \label{Zdir}
\end{equation}
This formula works well even for rather large $h$, when $h$ is
close to $R$.

More generally, the low-frequency impedance of the iris having 
a rectangular profile with an arbitrary aspect ratio is 
\begin{equation}
Z(\omega) = -i k Z_0 \frac{g h}{2\pi R} 
F \left ( \frac{h}{g} \right ) \ ,                 \label{irisg}
\end{equation}
where function F(x) is plotted in Fig.~6.

\begin{figure}
\centerline{\epsfig{figure=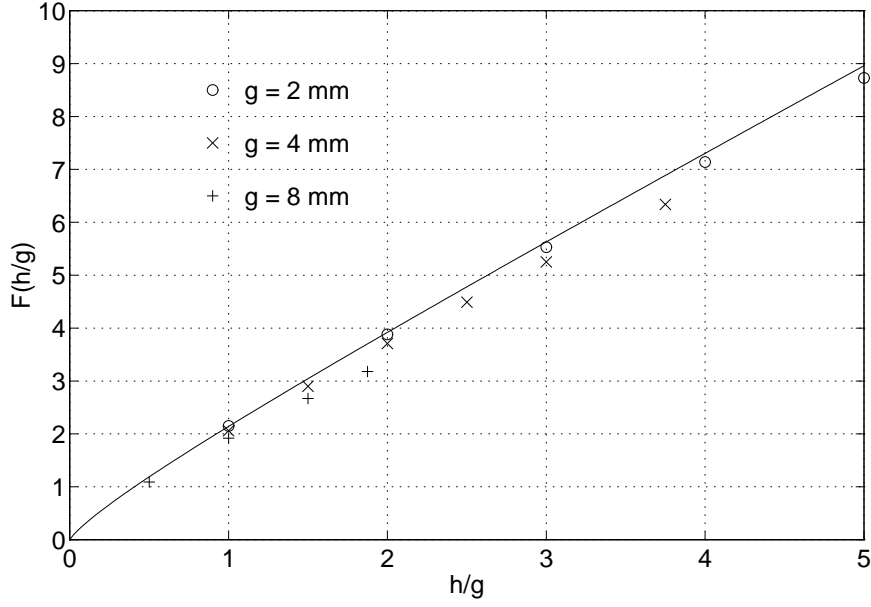,width=12cm}}
\caption{The inductance dependence on the aspect ratio of a 
rectangular iris, cf.\ Eq.~(\ref{irisg}). The marks show numerical
results for a beam pipe with radius $R=2$~cm, for comparison.} 
\end{figure}

The impedances of discontinuities having a triangle-shaped 
cross section with height (depth) $h$ and base $g$ along the beam 
are given below. When $g\ll h$, the low-frequency 
impedance of a triangular enlargement is 
\begin{equation}
Z(\omega) = -i k Z_0 \frac{1}{4\pi R}
\left (gh-\frac{g^2}{\pi}\right ) \ ,           \label{Ztre}
\end{equation}
and that of a triangular iris is
\begin{equation}
Z(\omega) = -i k Z_0 \frac{1}{4 R} \left [ h^2 + 
\frac{2gh}{\pi}(1 - \ln {2}) \right ] \ .        \label{Ztrir}
\end{equation}
For the case of shallow triangular perturbations, $h\ll g < R$,   
both the enlargement and contraction of the chamber have the same 
inductance, 
\begin{equation}
Z(\omega) = -i k Z_0 \frac{h^2 2\ln {2}}{\pi^2 R} \ ,         
                                                      \label{Zshtr}
\end{equation}
which is independent of $g$. 

\section*{Appendix}

\subsection*{Circular Chamber}

For a circular cross section of radius $b$ the eigenvalues 
$k_{nm}=\mu_{nm}/b$, where $\mu_{nm}$ is $m$th zero of the 
Bessel function $J_n(x)$, and the normalized EFs are 
\begin{equation} 
e_{nm}(r,\varphi) = \frac{J_n(k_{nm}r)}{\sqrt{N^E_{nm}}} 
    \left \{ \begin{array}{c} 
      \cos {n\varphi} \\ \sin {n\varphi} \end{array} 
    \right \}   \ ,                          \label{ecnm}
\end{equation} 
with $N^E_{nm} = \pi b^2 \epsilon_n J^2_{n+1}(\mu_{nm})/2$, 
where $\epsilon_0=2$ and $\epsilon_n=1$ for $n \ne 0$.
For TE-modes, $k'_{nm}=\mu'_{nm}/b$ with $J'_n(\mu'_{nm})=0$, 
and 
\begin{equation} 
h_{nm}(r,\varphi) = \frac{J_n(k'_{nm}r)}{\sqrt{N^H_{nm}}} 
    \left \{ \begin{array}{c} 
      \cos {n\varphi} \\ \sin {n\varphi} \end{array} 
    \right \}   \ ,                          \label{hcnm}
\end{equation} 
where $N^H_{nm} = \pi b^2 \epsilon_n (1-n^2/\mu'^2_{nm})
J^2_n(\mu'_{nm})/2$. In this case $\tilde{e}_\nu = 1/(2\pi b)$, 
which also follows from the Gauss law, and the formula  
for the inductive impedance takes an especially simple form, 
cf.~\cite{SK92,RLG}.

\subsection*{Rectangular Chamber}

For a rectangular chamber of width $a$ and height $b$ the 
eigenvalues are $k_{nm} = \pi \sqrt{n^2/a^2+m^2/b^2}$ with 
$n,m = 1,2, \ldots$, and the normalized EFs are 
\begin{equation} 
e_{nm}(x,y) = \frac{2}{\sqrt{ab}}  \sin {\frac{\pi n x}{a}} 
 \sin {\frac{\pi m y}{b}}  \ ,                  \label{ernm}
\end{equation} 
with $0 \le x \le a$ and $0 \le y \le b$. Let a hole be located 
in the side wall at $x=a, \ y=y_h$. From Eq.~(\ref{enorm}) 
after some algebra follows 
\begin{equation} 
 \tilde{e}_{\nu} = \frac{1}{b} \Sigma \left ( \frac{a}{b}, 
 \frac{y_h}{b} \right ) \ ,      \label{enrect}
\end{equation} 
where 
\begin{equation} 
 \Sigma (u,v) = \sum_{l=0}^{\infty} \frac { (-1)^l \sin 
 [\pi (2l+1) v ] }{ \cosh [\pi (2l+1) u / 2 ] }  \label{Sigma}
\end{equation} 
is a fast converging series; the behavior of $\Sigma (u,v)$ 
versus $v$ for different values of the aspect ratio $u$ is plotted 
in Ref.~\cite{SK92p}.

\section{Summary}

A review of calculating the beam coupling impedances of small
discontinuities was presented. We also collected some analytical 
formulas for the inductive contributions due to various small 
obstacles to the beam coupling impedances of the vacuum chamber. 

An importance of understanding these effects can be illustrated by 
the following example. An original design of the beam liner for the 
LHC vacuum chamber anticipated a circular liner with many circular
holes of 2-mm radius, providing the pumping area about 5\% of
the liner surface. Their total contribution to the low-frequency
coupling impedance was calculated \cite{SK92} to be 
$$ |Z/n| = 0.53 \; \Omega, \qquad |Z_\bot|=20 \; M\Omega /m \, ,$$
which was close or above (for $Z_\bot$) the estimated instability
threshold. A modified liner design had about the same pumping area
provided by rounded-end slots $1.5\times6$~mm${}^2$, which were
placed near the corners of the rounded-square cross section of the
liner \cite{SK95}. As a result of these changes, the coupling 
impedances were reduced by more than an order of magnitude, 30-50
times:
$$ |Z/n| = 0.017 \; \Omega, \qquad |Z_\bot|=0.4 \; M\Omega /m \, .$$ 
Now the pumping slots are not among the major contributors to the
impedance budget of the machine.

One should mention that these notes do not include more recent
developments, in particular, results for coaxial structures, 
frequency corrections for polarizabilities, etc. Some of these 
new results and proper references can be found in 
\cite{AccPhysHandbook}.

\end{document}